\documentclass[onecolumn]{revtex4}
\usepackage{graphicx}
\usepackage{amsmath}
\usepackage{amssymb}
\usepackage{amsthm}
\usepackage{bbm}
\usepackage[bookmarks = false,colorlinks = true,linkcolor = blue,urlcolor= black,citecolor = blue]{hyperref}
\begin{document}
\title{Quantum nonlocality without entanglement depending on nonzero prior probabilities in optimal unambiguous discrimination}
\author{Donghoon Ha}
\affiliation{Department of Applied Mathematics and Institute of Natural Sciences, Kyung Hee University, Yongin 17104, Republic of Korea}
\author{Jeong San Kim}
\email{freddie1@khu.ac.kr}
\affiliation{Department of Applied Mathematics and Institute of Natural Sciences, Kyung Hee University, Yongin 17104, Republic of Korea}
\begin{abstract}
Nonlocality without entanglement(NLWE) is a nonlocal phenomenon that occurs in quantum state discrimination of multipartite separable states.
In the discrimination of orthogonal separable states, the term NLWE is used when the quantum states cannot be discriminated perfectly by local operations and classical communication.
In this case,
the occurrence of NLWE is independent of nonzero prior probabilities of quantum states
being prepared.
Recently, it has been found that the occurrence of NLWE can depend on nonzero prior probabilities in minimum-error discrimination of nonorthogonal separable states.
Here, we show that even in optimal unambiguous discrimination, the occurrence of NLWE can depend on nonzero prior probabilities.
We further show that NLWE can occur regardless of nonzero prior probabilities, even if only one state can be locally discriminated without error.
Our results provide new insights into classifying sets of multipartite quantum states in terms of quantum state discrimination.
\end{abstract}
\maketitle
%%%%%%%%%%%%%%%%(SECTION)
\section*{Introduction}
\indent Although orthogonal quantum states can be perfectly discriminated in quantum mechanics, 
there also exists a set of orthogonal, non-entangled(separable) states in multipartite quantum systems
that cannot be perfectly discriminated only by local operations and classical communications(LOCC)\cite{chef2000,barn2009,berg2010,bae2015,benn19991}.
This set of multipartite quantum states reveals the possible existence of separable measurement(SEP) that cannot be realized by LOCC. In discriminating separable states, a phenomenon that can be achieved by global measurements but cannot be achieved only by LOCC is called {\em nonlocality without entanglement}(NLWE).\\
\indent The first NLWE phenomenon was shown through nine $3\otimes3$ orthogonal product states\cite{benn19991}.
On the other hand, Walgate et al.\cite{walg2000} showed that any two multipartite orthogonal pure states can be perfectly discriminiated within finite-round using LOCC. Since then, there have been many studies focused on local indistinguishability or local distinguishability of multipartite orthogonal states\,\cite{benn19992,ghos2001,walg2002,divi2003,ghos2004,fan2004,watr2005,nise2006,cohe2007,duan2009,klei2011,band2012,yang2013,zhan20142,chit20141,wang2015,yang2016,zhan2016,zhan2017,hald2019,jian2020}. Recently, there has also been an effort to find NLWE in generalized probabilistic theories\,\cite{bhat2020}.\\
\indent The NLWE phenomenon occurs in the perfect discrimination of orthogonal separable states as well as minimum-error discrimination (MD)\cite{hels1976,hole1979,yuen1975} of nonorthogonal separable states. 
MD of two multipartite nonorthogonal pure states can always be achieved using LOCC\cite{virm2001}, 
regardless of the prior probabilities. However, MD of more than two states is not
(e.g., double trine ensemble\cite{pere1991,chit2013});
the minimum error probability cannot be obtained using asymptotic LOCC\cite{chit20142,alocc}.
Recently, Ha and Kwon\cite{ha2021} found a set of three nonorthogonal $2\otimes2$ product states
where NLWE in terms of MD occurs depending on the nonzero prior probabilities.
In perfect discrimination of orthogonal states, the nonzero prior probabilities have no effect on the local indistinguishability or the local distinguishability.
That is, NLWE in terms of perfect discrimination does not depend on nonzero prior probabilities.\\
\indent In unambiguous discrimination(UD)\cite{ivan1987,pere1988,diek1988,jaeg1995},
quantum states are discriminated without error (i.e., unambiguously), instead, allowing inconclusive results. 
In order for pure states to be unambiguously discriminated, they must be linearly independent\cite{chef19981}. 
Moreover, in order to unambiguously discriminate a multipartite pure state using LOCC, the existence of a product state that detects only the state is additionally required\cite{chef2004,duan2007,band2009}.\\
\indent The optimization of UD is to minimize the probability of obtaining inconclusive results.
Whereas optimal UD of two multipartite pure states can always be achieved using LOCC\cite{virm2001,chen2001,chen2002,ji2005}, optimal UD of more than two multipartite states is not in general.
A typical example is ``indistinguishable product basis'' presented by Duan et al.\cite{duan2007},
in which all states cannot be unambiguously discriminated using LOCC.
We also note that NLWE in terms of optimal UD may occur regardless of nonzero prior probabilities.
However, it is still a natural question whether there is a set of separable states that cause NLWE in terms of optimal UD depending on nonzero prior probabilities.\\
\indent In this work, we consider a complete product basis, in which only one state can be discriminated unambiguously using LOCC, and establish a necessary and sufficient condition that the occurrence of NLWE in terms of optimal UD depends on nonzero prior probabilities (Corollary 2).
We provide an example of six $2\otimes3$ product states where NLWE in terms of optimal UD occurs depending on nonzero prior probabilities (Example 1). We also provide an example that is not indistinguishable product basis
but causes NLWE in terms of optimal UD regardless of nonzero prior probabilities (Example 2).
We further investigate the case where optimal UD measurement detects only one state unambiguously (Theorem 1), and obtain a necessary and sufficient condition for optimal UD measurement to detect two or more states unambiguously, regardless of nonzero prior probabilities (Corollary 1).

%%%%%%%%%%%%%%%%(SECTION)
\section*{Results}
%%%%%%%%%%%%%%%%(SUBSECTION)
\subsection*{Optimal Unambiguous Discrimination}
 Let $\{|\psi_{1}\rangle,\ldots,|\psi_{n}\rangle\}$ be a set of $n$ linearly independent pure states 
and $\eta_{i}$ be the corresponding non-zero probability of $|\psi_{i}\rangle$ being prepared for $i=1,\ldots,n$.
The UD of $\{\eta_{i},|\psi_{i}\rangle\}_{i=1}^{n}$ is to discriminate the prepared state without error using 
 a positive operator-valued measure(POVM) that consists of $n+1$ positive semidefinite operators $M_{0},M_{1},\ldots,M_{n}$ on $\mathcal{H}_{\rm S}$ satisfying $\sum_{i=0}^{n}M_{i}=\mathbbm{1}$. Here $\mathcal{H}_{\rm S}$ is the subspace spanned by $\{|\psi_{1}\rangle,\ldots,|\psi_{n}\rangle\}$ and $\mathbbm{1}$ is the identity operator on $\mathcal{H}_{\rm S}$. $M_{0}$ provides inconclusive results for the prepared state and $M_{i}(i=1,\ldots,n)$ gives conclusive results for $|\psi_{i}\rangle$ unambiguously.\\
\indent The no-error condition 
for $\{M_{i}\}_{i=0}^{N}$ to discriminate $\{|\psi_{i}\rangle\}_{i=1}^{n}$
unambiguously\cite{pang2009}, that is, $\langle\psi_{i}|M_{j}|\psi_{i}\rangle=0$ $\forall i\neq j$, requires
\begin{equation}
M_{i}=p_{i} |\tilde{\psi}_{i}\rangle\langle\tilde{\psi}_{i}|,\ i=1,\ldots,n,
\end{equation}
where $p_{i}$ is the probability of success 
in unambiguously discriminating $|\psi_{i}\rangle$ and 
$|\tilde{\psi}_{i}\rangle$ is the reciprocal vector defined by 
\begin{equation}
|\tilde{\psi}_{i}\rangle=\sum_{j=1}^{n}(\Omega^{-1})_{ji}|\psi_{j}\rangle,
\end{equation}
where $\Omega^{-1}$ is the inverse matrix of $n$-by-$n$ Gram matrix $\Omega$
with $(\Omega)_{ij}=\langle\psi_{i}|\psi_{j}\rangle$.
We note that $\langle\tilde{\psi}_{i}|\tilde{\psi}_{j}\rangle=(\Omega^{-1})_{ij}$.
Because $\{|\tilde{\psi}_{1}\rangle,\ldots,|\tilde{\psi}_{n}\rangle\}$ is a set of linearly independent vectors in $\mathcal{H}_{\rm S}$ satisfying $\langle\tilde{\psi}_{i}|\psi_{j}\rangle=\delta_{ij}$,
selecting a UD measurement can be 
understood as selecting a point $\vec{p}=(p_{1},\ldots,p_{n})^{\sf T}\in\mathbb{R}^{n}$ such that
\begin{equation}
\sum_{i=1}^{n}p_{i}|\tilde{\psi}_{i}\rangle\langle\tilde{\psi}_{i}|\geqslant0,\quad
\mathbbm{1}-\sum_{i=1}^{n}p_{i}|\tilde{\psi}_{i}\rangle\langle\tilde{\psi}_{i}|\geqslant0.
\end{equation}
\indent The optimization task of UD is to maximize the average probability of success, 
\begin{equation}
P=\sum_{i=1}^{n}\eta_{i}p_{i}.
\end{equation}
whose maximum is denoted by $P_{\max}$.
The optimal $\vec{p}$ that gives $P_{\max}$ is unique\cite{sugi2010}, which can also be verified from optimality condition including Lagrangian stability; see Lemma 1 in the Methods Section.\\
\indent 
Some components of optimal $\vec{p}$ may be zero.
In other words, the optimal UD measurement of $\{\eta_{i},|\psi_{i}\rangle\}_{i=1}^{n}$ may not detect all given quantum states unambiguously.
If $p_{i}$ is zero, the UD measurement does not detect $|\psi_{i}\rangle$ unambiguously.
The following theorem shows a necessary and sufficient condition for the optimal UD measurement of $\{\eta_{i},|\psi_{i}\rangle\}_{i=1}^{n}$ to detect only $|\psi_{1}\rangle$ unambiguously.
The proof of Theorem 1 is provided in the Methods Section.\\
\indent{\bf Theorem 1.} For the optimal UD of $\{\eta_{i},|\psi_{i}\rangle\}_{i=1}^{n}$, the optimal measurement detects
only $|\psi_{1}\rangle$ unambiguously if and only if
\begin{equation}\label{eq:conofprior}
\frac{\eta_{i}}{\eta_{1}}
\leqslant
\frac{|\langle\tilde{\psi}_{i}|\tilde{\psi}_{1}\rangle|^{2}}{|\langle\tilde{\psi}_{1}|\tilde{\psi}_{1}\rangle|^{2}}\ \forall i.
\end{equation}
\indent In the case of $n=2$\cite{jaeg1995}, 
it is easy to see that the condition $\eta_{2}/\eta_{1}\leqslant|\langle\psi_{1}|\psi_{2}\rangle|^{2}$ is equivalent that 
the optimal $p_{2}$ is zero.
As we can check in the proof of Theorem 1 in Method Section, the choice of $|\psi_{1}\rangle$ in Theorem 1 can be arbitrary. That is, any of $\{|\psi_{1}\rangle,\ldots,|\psi_{n}\rangle\}$ can be used to play the role of $|\psi_{1}\rangle$ in Theorem 1.\\
\indent{\bf Corollary 1.}
For any set of nonzero prior probabilities $\{\eta_{i}\}_{i=1}^{n}$,
the optimal UD measurement of $\{\eta_{i},|\psi_{i}\rangle\}_{i=1}^{n}$
detects more than one state unambiguously
if and only if, for any $j$, there is $k$ satisfying $\langle\tilde{\psi}_{j}|\tilde{\psi}_{k}\rangle=0$.
\begin{proof}
For fixed $j$, let us first suppose that
there exists $k$ satisfying $\langle\tilde{\psi}_{j}|\tilde{\psi}_{k}\rangle=0$, which implies
\begin{equation}\label{eq:ekejtpk}
\frac{\eta_{k}}{\eta_{j}}
>
\frac{|\langle\tilde{\psi}_{k}|\tilde{\psi}_{j}\rangle|^{2}}{|\langle\tilde{\psi}_{j}|\tilde{\psi}_{j}\rangle|^{2}}
\end{equation} 
because the left-hand side is positive and the right-hand side is zero. 
Thus, from Theorem 1, the optimal UD measurement of $\{\eta_{i},|\psi_{i}\rangle\}_{i=1}^{n}$
cannot be achieved by detecting only $|\psi_{j}\rangle$ unambiguously.
Since this argument is true for any $|\psi_{j}\rangle$, 
the optimal UD measurement must detect more than one state unambiguously.\\
\indent Now, let us suppose that $\langle\tilde{\psi}_{j}|\tilde{\psi}_{k}\rangle\neq0$ for all $k$.
We can see that 
\begin{equation}\label{eq:ekejtpk}
\frac{\eta_{k}}{\eta_{j}}
=
\frac{|\langle\tilde{\psi}_{k}|\tilde{\psi}_{j}\rangle|^{2}}{|\langle\tilde{\psi}_{j}|\tilde{\psi}_{j}\rangle|^{2}}\ \forall k
\end{equation} 
with nonzero prior probabilities defined as 
\begin{equation}
\eta_{i}=\frac{|\langle\tilde{\psi}_{i}|\tilde{\psi}_{j}\rangle|^{2}}{\sum_{i'=1}^{n}|\langle\tilde{\psi}_{i'}|\tilde{\psi}_{j}\rangle|^{2}}.
\end{equation}
Thus, from Theorem 1, the optimal UD measurement of
$\{\eta_{i},|\psi_{i}\rangle\}_{i=1}^{n}$
detects only $|\psi_{j}\rangle$ unambiguously.
\end{proof}
\indent For example, let us consider four pure states
\begin{equation}\label{eq:exnpdo}
\begin{array}{rcl}
|\psi_{1}\rangle&=&\frac{1}{\sqrt{3}}(|0\rangle-|2\rangle-|3\rangle),\\[1mm]
|\psi_{2}\rangle&=&\frac{1}{\sqrt{3}}(|1\rangle-|2\rangle+|3\rangle),\\[1mm]
|\psi_{3}\rangle&=&|2\rangle,\\[1mm]
|\psi_{4}\rangle&=&|3\rangle,
\end{array}
\end{equation}
with the reciprocal vectors
\begin{equation}
\begin{array}{rcl}
|\tilde{\psi}_{1}\rangle&=&\sqrt{3}|0\rangle,\\[1mm]
|\tilde{\psi}_{2}\rangle&=&\sqrt{3}|1\rangle,\\[1mm]
|\tilde{\psi}_{3}\rangle&=&|0\rangle+|1\rangle+|2\rangle,\\[1mm]
|\tilde{\psi}_{4}\rangle&=&|0\rangle-|1\rangle+|3\rangle.
\end{array}
\end{equation}
Since $\langle\tilde{\psi}_{1}|\tilde{\psi}_{2}\rangle$ and $\langle\tilde{\psi}_{3}|\tilde{\psi}_{4}\rangle$ are zero, 
for any $i$, there exists $j$ satisfying $\langle\tilde{\psi}_{i}|\tilde{\psi}_{j}\rangle=0$.
Therefore, for any set of nonzero prior probabilities, 
the optimal UD measurement detects more than one state unambiguously.
%%%%%%%%%%%%%%%%(SUBSECTION)
\subsection*{Multipartite State Discrimination}
\indent Let $\mathcal{H}$ be a multipartite Hilbert space 
and $\{|\Psi_{i}\rangle\}_{i=1}^{n}$ a set of $n$ linearly independent pure states
in $\mathcal{H}$,
where each $|\Psi_{i}\rangle$ is prepared with nonzero prior probability $\xi_{i}$.
We use $|\Psi_{i}\rangle$, $|\tilde{\Psi}_{i}\rangle$, and $P_{\max}^{\mbox{\tiny Global}}$ instead of $|\psi_{i}\rangle$, $|\tilde{\psi}_{i}\rangle$, and $P_{\max}$ to highlight the multipartite quantum system to be considered.
Also, we use $P_{\rm max}^{\overline{\mbox{\tiny LOCC}}}$ to denote the maximum of average success probability
that can be obtained in UD restricted to asymptotic LOCC,
whereas $P_{\max}^{\mbox{\tiny Global}}$ is
the maximum of average success probability
that can be obtained using global measurement.
If there exists a product state $|\Phi\rangle$ that satisfies $\langle\Phi|\Psi_{1}\rangle\neq0$ and $\langle\Phi|\Psi_{i}\rangle=0\,\forall i\neq 1$, $|\Psi_{1}\rangle$ can be unambiguously discriminated using finite-round LOCC, 
otherwise the state cannot be unambiguously discriminated 
even using SEP\,\cite{chef2004,duan2007,band2009}.
In particular, when $\{|\Psi_{i}\rangle\}_{i=1}^{n}$ is a basis of $\mathcal{H}$ (that is, $\mathcal{H}_{\rm S}=\mathcal{H}$), 
the existence of a product state $|\Phi\rangle$ satisfying $\langle\Phi|\Psi_{1}\rangle\neq0$ and $\langle\Phi|\Psi_{i}\rangle=0\,\forall i\neq 1$ is determined by the corresponding reciprocal vector $|\tilde{\Psi}_{1}\rangle$\cite{duan2007}.
To be precise, $|\Psi_{1}\rangle$ can be unambiguously discriminated using LOCC
if and only if $|\tilde{\Psi}_{1}\rangle$ is a product vector. 
Note that when $\mathcal{H}_{\rm S}\neq\mathcal{H}$, even if $|\tilde{\Psi}_{1}\rangle$ is entangled, 
a product state that detects only $|\Psi_{1}\rangle$ may exist in $\mathcal{H}$.\\
\indent In the case of $n=2$\,\cite{virm2001,chen2001,chen2002,ji2005}, 
the globally optimal UD is always possible only with finite-round LOCC 
no matter what two pure states with arbitrary prior probabilities are given, i.e.,
$P_{\rm max}^{\overline{\mbox{\tiny LOCC}}}=P_{\max}^{\mbox{\tiny Global}}$.
However, in the case of $n>2$, $P_{\rm max}^{\overline{\mbox{\tiny LOCC}}}$ can be less than $P_{\rm max}^{\mbox{\tiny Global}}$.
An example is the indistinguishable product basis\cite{duan2007} in which 
no state can be unambiguously discriminated using asymptotic LOCC, 
i.e., $P_{\max}^{\overline{\mbox{\tiny LOCC}}}=0$.
It should be noted that $P_{\max}^{\overline{\mbox{\tiny LOCC}}}=0$ means $P_{\max}^{\overline{\mbox{\tiny LOCC}}}<P_{\max}^{\mbox{\tiny Global}}$ because
$P_{\max}^{\mbox{\tiny Global}}$ is always nonzero for any linearly independent pure states. 
Moreover, this example illustrates the case that  
NLWE in terms of optimal UD occurs regardless of nonzero prior probabilities.\\
\indent The following theorem shows that if $\{|\Psi_{i}\rangle\}_{i=1}^{n}$ is a basis of $\mathcal{H}$
and only $|\Psi_{1}\rangle$ can be unambiguously discriminated using LOCC, 
$P_{\max}^{\overline{\mbox{\tiny LOCC}}}=P_{\max}^{\mbox{\tiny Global}}$
is determined by whether the optimal UD measurement requires unambiguous detection other than $|\Psi_{1}\rangle$.
The proof of Theorem 2 is provided in the Methods Section.\\
\indent{\bf Theorem 2.}
When $\mathcal{H}_{\sf S}=\mathcal{H}$ and 
 only $|\tilde{\Psi}_{1}\rangle$ is a product vector,
the globally optimal UD of $\{\xi_{i},|\Psi_{i}\rangle\}_{i=1}^{n}$ 
can be achieved using LOCC if and only if 
\begin{equation}\label{eq:onlyonecon}
\frac{\xi_{i}}{\xi_{1}}
\leqslant
\frac{|\langle\tilde{\Psi}_{i}|\tilde{\Psi}_{1}\rangle|^{2}}{
|\langle\tilde{\Psi}_{1}|\tilde{\Psi}_{1}\rangle|^{2}}\ \ \forall i.
\end{equation}
\indent Theorem 2 tells us that the possibility of globally optimal UD using LOCC 
can depend on nonzero prior probabilities.
In other words, when $|\Psi_{1}\rangle,\ldots,|\Psi_{n}\rangle$ are fixed,
both $P_{\max}^{\overline{\mbox{\tiny LOCC}}}=P_{\max}^{\mbox{\tiny Global}}$ and $P_{\max}^{\overline{\mbox{\tiny LOCC}}}<P_{\max}^{\mbox{\tiny Global}}$ can occur 
according to $\xi_{1},\ldots,\xi_{n}$.
 As we can check in the proof of Theorem 2 in Method Section, the choice of $|\tilde{\Psi}_{1}\rangle$ in Theorem 2 can be arbitrary. That is, any of $\{|\tilde{\Psi}_{1}\rangle,\ldots,|\tilde{\Psi}_{n}\rangle\}$ can be used to play the role of $|\tilde{\Psi}_{1}\rangle$ in Theorem 2.\\
\indent{\bf Corollary 2.} For the optimal UD of $n$ product states $|\Psi_{1}\rangle,\ldots,|\Psi_{n}\rangle$ such that
$\mathcal{H}_{\sf S}=\mathcal{H}$ and only $|\tilde{\Psi}_{1}\rangle$ is a product vector,
if $\langle\tilde{\Psi}_{i}|\tilde{\Psi}_{1}\rangle$ is nonzero for all $i$,
NLWE occurs depending on the nonzero prior probabilities.
Otherwise NLWE occurs for any nonzero prior probabilities.
\begin{proof}
If $\langle\tilde{\Psi}_{i}|\tilde{\Psi}_{1}\rangle\neq0$ for all $i$, satisfying Inequality \eqref{eq:onlyonecon} for all $i$ depends on nonzero prior probabilities;
Inequality \eqref{eq:onlyonecon} for each $i$ holds by the nonzero prior probabilities defined as
\begin{equation}
\xi_{i}=\frac{|\langle\tilde{\Psi}_{i}|\tilde{\Psi}_{1}\rangle|^{2}}{\sum_{j=1}^{n}|\langle\tilde{\Psi}_{j}|\tilde{\Psi}_{1}\rangle|^{2}}, 
\end{equation}
whereas Inequality \eqref{eq:onlyonecon} for $i=2$ does not hold by the nonzero prior probabilities defined as
\begin{equation}
\xi_{i}=\left\{
\begin{array}{lll}
\frac{1+|\langle\tilde{\Psi}_{2}|\tilde{\Psi}_{1}\rangle|^{2}}{1+\sum_{j=1}^{n}|\langle\tilde{\Psi}_{j}|\tilde{\Psi}_{1}\rangle|^{2}}&\mbox{for}&i=2,\\[2mm]
\frac{|\langle\tilde{\Psi}_{i}|\tilde{\Psi}_{1}\rangle|^{2}}{1+\sum_{j=1}^{n}|\langle\tilde{\Psi}_{j}|\tilde{\Psi}_{1}\rangle|^{2}}&\mbox{for}&i\neq2.
\end{array}
\right.
\end{equation}
It follows from Theorem 2 that the possibility of the globally optimal UD using LOCC depends on nonzero prior probabilities.
Thus, NLWE occurs depending on the nonzero prior probabilities.\\
\indent However, if $\langle\tilde{\Psi}_{j}|\tilde{\Psi}_{1}\rangle=0$ for some $j$,
no set of nonzero prior probabilities satisfies
Inequality \eqref{eq:onlyonecon} for all $i$.
It follows from Theorem 2 that for any set of nonzero prior probabilities $\{\xi_{i}\}_{i=1}^{n}$, the globally optimal UD of $\{\xi_{i},|\Psi_{i}\rangle\}_{i=1}^{n}$ cannot be achieved using LOCC.
Therefore, NLWE occurs regardless of nonzero prior probabilities.
\end{proof}
\indent{\bf Example 1.} Let us consider six $2\otimes3$ product states,
\begin{equation}\label{eq:sixex} 
\begin{array}{rclrcl}
|\Psi_{1}\rangle&=&\frac{1}{6}(|0\rangle+|1\rangle)
\otimes(4|0\rangle+|1\rangle+|2\rangle),\\[1mm]
|\Psi_{2}\rangle&=&\frac{1}{3}|1\rangle\otimes(2|0\rangle+2|1\rangle+|2\rangle),\\[1mm]
|\Psi_{3}\rangle&=&\frac{1}{3}|1\rangle\otimes(2|0\rangle+|1\rangle+2|2\rangle),\\[1mm]
|\Psi_{4}\rangle&=&\frac{1}{5}(2|0\rangle+|1\rangle)\otimes(2|1\rangle+|2\rangle),\\[1mm]
|\Psi_{5}\rangle&=&\frac{1}{5}(|0\rangle+2|1\rangle)\otimes(|1\rangle+2|2\rangle),\\[1mm]
|\Psi_{6}\rangle&=&\frac{1}{2}(|0\rangle+|1\rangle)\otimes(|1\rangle+|2\rangle).
\end{array}
\end{equation}
Example 1 satisfies $\mathcal{H}_{\rm S}=\mathcal{H}$ because
\begin{eqnarray}\label{eq:hshc}
\begin{array}{l}
{\rm span}\{|\Psi_{1}\rangle,|01\rangle,|02\rangle,|10\rangle,|11\rangle,|12\rangle\}=\mathcal{H},\\[1mm]
{\rm span}\{|\Psi_{2}\rangle,|\Psi_{3}\rangle,|12\rangle\,\}=
{\rm span}\{|10\rangle,|11\rangle,|12\rangle\},\\[1mm]
{\rm span}\{|\Psi_{4}\rangle,|\Psi_{5}\rangle,|\Psi_{6}\rangle\}=
{\rm span}\{|01\rangle,|02\rangle+|11\rangle,|12\rangle\}.
\end{array}
\end{eqnarray}
The reciprocal vectors are
\begin{equation}\begin{array}{rcl}
|\tilde{\Psi}_{1}\rangle&=&\frac{3}{2}|00\rangle,\\[1mm]
|\tilde{\Psi}_{2}\rangle&=&\frac{3}{2}(|00\rangle-2|02\rangle-|10\rangle+2|11\rangle),\\[1mm]
|\tilde{\Psi}_{3}\rangle&=&-3(|00\rangle-|02\rangle-|10\rangle+|11\rangle),\\[1mm]
|\tilde{\Psi}_{4}\rangle&=&\frac{5}{6}(3|00\rangle+4|01\rangle-8|02\rangle
-3|10\rangle+2|11\rangle+2|12\rangle),\\[1mm]
|\tilde{\Psi}_{5}\rangle&=&\frac{5}{3}(3|00\rangle+|01\rangle-5|02\rangle
-3|10\rangle+2|11\rangle+2|12\rangle),\\[1mm]
|\tilde{\Psi}_{6}\rangle&=&-\frac{1}{2}(13|00\rangle+8|01\rangle-28|02\rangle
-12|10\rangle+8|11\rangle+8|12\rangle).
\end{array}\end{equation}
Only $|\tilde{\Psi}_{1}\rangle$ is a product vector
because the reduced non-negative operators of $|\tilde{\Psi}_{2}\rangle\langle\tilde{\Psi}_{2}|,\ldots,|\tilde{\Psi}_{6}\rangle\langle\tilde{\Psi}_{6}|$ for the qubit 
are not all rank one. \\
\begin{figure}[!t]
\centerline{\includegraphics*[bb=30 20 430 310,scale=0.8]{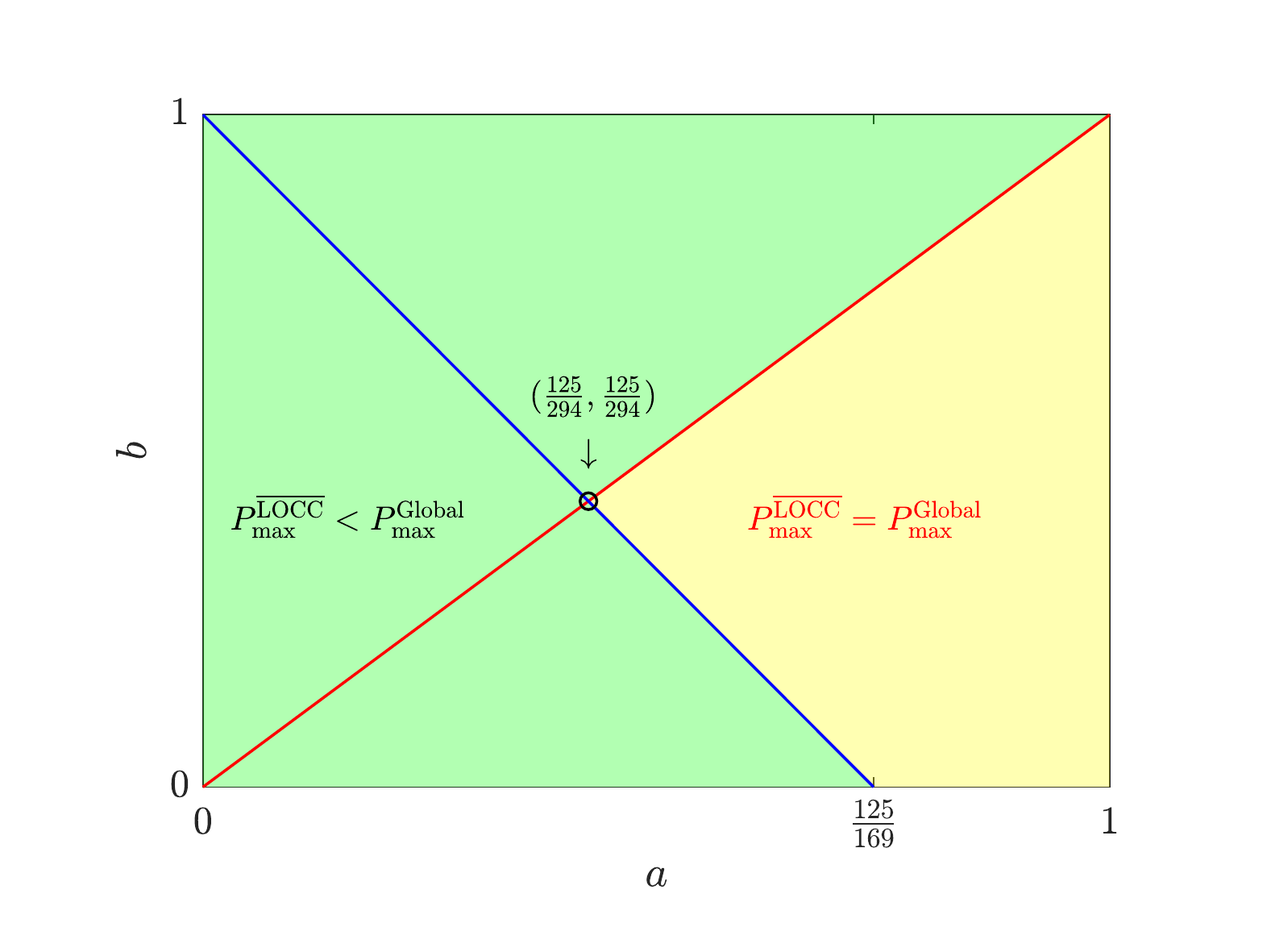}}
\caption{{\bf The occurrence of NLWE according to $a,b\in(0,1)$.} 
NLWE occurs in the green region where $a$ is smaller than $b$ or $\frac{125}{169}(1-b)$.
However, NLWE does not occur in the yellow area where $a$ is equal to or greater than $b$ and $\frac{125}{169}(1-b)$.}\label{fig:nlwe}
\end{figure}
\indent According to Theorem 2, we can achieve 
the globally optimal UD between six states of \eqref{eq:sixex} using LOCC if and only if
\begin{equation}
\frac{\xi_{1}}{9}\geqslant\frac{\xi_{2}}{9},\frac{\xi_{3}}{36},\frac{\xi_{4}}{25},\frac{\xi_{5}}{100},\frac{\xi_{6}}{169}.
\end{equation}
Furthermore, since the given six states are product vectors, 
we can see from Corollary 2 that the occurrence of NLWE 
in terms of optimal UD depends on nonzero prior probabilities.
Figure \ref{fig:nlwe} illustrates two cases classified by the occurrence of NLWE 
when considering nonzero prior probabilities determined by 
two variables $a\in(0,1)$ and $b\in(0,1)$ as follows:
\begin{equation}
\xi_{1}=\xi_{2}=\frac{9a}{143},\ \xi_{3}=\frac{18(1-a)}{143},\ 
\xi_{4}=\frac{25b}{143},\ \xi_{5}=\frac{100b}{143},\ \xi_{6}=\frac{125(1-b)}{143}.
\end{equation}
For this case, NLWE does not occur if and only if
$a$ is equal to or greater than $b$ and $\frac{125}{169}(1-b)$.\\
\indent {\bf Example 2.} Let us consider the case where the first state of Example 1 is replaced with 
\begin{equation}\begin{array}{rcl}
|\Psi_{1}\rangle&=&\frac{1}{15}(3|0\rangle+4|1\rangle)\otimes(|0\rangle+2|1\rangle+2|2\rangle).
\end{array}\end{equation}
In this case as well, Eq.\,\eqref{eq:hshc} is satisfied, 
so the given six states form a basis for $\mathcal{H}$.
The reciprocal set corresponding to $\{|\Psi_{i}\rangle\}_{i=1}^{6}$ consists of 
\begin{equation}\begin{array}{rcl}
|\tilde{\Psi}_{1}\rangle&=&5|00\rangle,\\[1mm] 
|\tilde{\Psi}_{2}\rangle&=&-\frac{3}{2}(2|02\rangle+|10\rangle-2|11\rangle),\\[1mm]
|\tilde{\Psi}_{3}\rangle&=&-2|00\rangle+3|02\rangle+3|10\rangle-3|11\rangle,\\[1mm]
|\tilde{\Psi}_{4}\rangle&=&\frac{5}{18}(4|00\rangle+12|01\rangle-24|02\rangle
-9|10\rangle+6|11\rangle+6|12\rangle),\\[1mm]
|\tilde{\Psi}_{5}\rangle&=&\frac{5}{9}(4|00\rangle+3|01\rangle-15|02\rangle
-9|10\rangle+6|11\rangle+6|12\rangle),\\[1mm]
|\tilde{\Psi}_{6}\rangle&=&-\frac{2}{3}(10|00\rangle+6|01\rangle-21|02\rangle
-9|10\rangle+6|11\rangle+6|12\rangle).
\end{array}\end{equation}
Also, only $|\tilde{\Psi}_{1}\rangle$ is a product vector because
the reduced nonnegative operator of $|\tilde{\Psi}_{i}\rangle\langle\tilde{\Psi}_{i}|$ is not rank one for $i\neq1$.
Since $\langle\tilde{\Psi}_{1}|\tilde{\Psi}_{2}\rangle$ is zero,
according to Corollary 2, no matter what nonzero prior probabilities $\xi_{1},\ldots,\xi_{6}$ are given,
the globally optimal UD of $\{\xi_{i},|\Psi_{i}\rangle\}_{i=1}^{6}$ cannot be achieved using LOCC.
This example that is not an indistinguishable product basis\cite{duan2007}
shows that NLWE can occur regardless of nonzero prior probabilities.
We also note that, in two qubits, there is no basis consisting of 
four product states that satisfy the assumption of Theorem 2;
see Methods Section.
%%%%%%%%%%%%%%%%(SECTION)
\section*{Discussion}
We have shown that NLWE in terms of optimal UD can depend on the nonzero prior probabilities of  quantum states being prepared. 
We have first established a necessary and sufficient condition for optimal UD measurement of linearly independent pure states to detect only one specific state unambiguously (Theorem 1).
We have also provided a necessary and sufficient condition for optimal UD measurement to detect two or more states unambiguously, regardless of nonzero prior probabilities (Corollary 1).
By generalizing Theorem 1 to multipartite state discrimination, we have established a necessary and sufficient condition for globally optimal UD of linearly independent pure states using LOCC provided that only one state can be discriminated unambiguously (Theorem 2).
From Theorem 2, we have provided a sufficient condition that NLWE occurs depending on nonzero prior probabilities, as well as a sufficient condition that NLWE occurs regardless of nonzero prior probabilities (Corollary 2).
Finally, we have illustrated the occurrence of NLWE in Corollary 2 by providing Examples 1 and 2.\\
\indent Our result means that sets of linearly independent product states can be classified into three types in terms of optimal UD: Type I where NLWE does not occur regardless of nonzero prior probabilities (e.g., two multipartite pure states\cite{walg2000,virm2001,chen2001,chen2002,ji2005}), Type II where NLWE occurs regardless of nonzero prior probabilities (e.g., domino states\cite{benn19991}, indistinguishable product basis\cite{duan2007}, and Example 2), and Type III where NLWE occurs depending on nonzero prior probabilities (e.g. Example 1).
Surprisingly, Type III is a new type that only occurs in nonorthogonal states.\\
\indent The quantum states of Type III can be useful for possible construction of cryptographical schemes such as quantum data hiding\cite{terh2001,divi2002,egge2002,matt2009} or quantum secret sharing\cite{raha2015,wang2017} with nonorthogonal states. In most quantum data-hiding schemes, classical bits are hidden by orthogonal quantum states of Type II which cannot be perfectly discriminated using LOCC regardless of nonzero prior probabilities. In many quantum secret sharing schemes, classical bits are shared among parties using orthogonal quantum states of Type I that can be perfectly discriminated using LOCC regardless of nonzero prior probabilities. By using nonorthogonal quantum states of Type III, an unified cryptographic scheme can be constructed that can be both data-hiding and secret-sharing scheme depending on nonzero prior probabilities. We believe this is an interesting challenge as a future work.\\
\indent Also, we have provided a new example of Type II (Example 2).
Interestingly, this example is different from the domino states\cite{benn19991},
which can be unambiguously discriminated using LOCC.
Moreover, this example is also different from the indistinguishable product basis\cite{duan2007},
which cannot be unambiguously discriminated using LOCC.
In Example 2, one state can be unambiguously discriminated using LOCC, but all other states cannot.
This suggests that Type II can be subdivided into more inequivalent classes.
%%%%%%%%%%%%%%%%(SECTION)
\section*{Methods}
%%%%%%%%%%%%%%%%(SUBSECTION)
\subsection*{Semidefinite programming to optimal UD of linearly independent pure states}
\indent The optimal UD of $\{\eta_{i},|\psi_{i}\rangle\}_{i=1}^{n}$ can be expressed as 
\begin{equation}\label{eq:primal}
\begin{array}{lll}
\mbox{maximize}& {\rm Tr}(\rho M_{\vec{p}})\\[1mm]
\mbox{subject to}&M_{\vec{p}}\geqslant 0
\ \mbox{and}\ 
\mathbbm{1}-M_{\vec{p}}\geqslant 0,
\end{array}
\end{equation}
where 
\begin{equation}\label{eq:rhomp}
\rho=\sum_{i=1}^{n}\eta_{i}|\psi_{i}\rangle\langle\psi_{i}|,
\ \
M_{\vec{p}}=\sum_{i=1}^{n}p_{i}|\tilde{\psi}_{i}\rangle\langle\tilde{\psi}_{i}|.
\end{equation}
Eq. \eqref{eq:primal} is called a primal problem.
$M_{\vec{p}}$ is said to be feasible if it safisfies the two constraints of \eqref{eq:primal},
and strictly feasible if both $M_{\vec{p}}$ and $\mathbbm{1}-M_{\vec{p}}$ are positive-definite.\\
\indent When constructing the Lagrangian\cite{boyd2004} as 
\begin{eqnarray}
\mathcal{L}(M_{\vec{p}},E,K)=
{\rm Tr}(\rho M_{\vec{p}})
+{\rm Tr}(M_{\vec{p}}E)+{\rm Tr}[(\mathbbm{1}-M_{\vec{p}})K]
={\rm Tr}K+{\rm Tr}[M_{\vec{p}}(\rho+E-K)],
\end{eqnarray}
we can derive the dual problem\cite{elda2003,elda2004} of \eqref{eq:primal}:
\begin{equation}\label{eq:dual}
\begin{array}{lll}
\mbox{minimize}& {\rm Tr}K\\[1mm]
\mbox{subject to}&
\langle\tilde{\psi}_{i}|E|\tilde{\psi}_{i}\rangle\geqslant 0\ \forall i,\ K\geqslant 0,\ \mbox{and}\ K=\rho+E,
\end{array}
\end{equation}
where $E$ and $K$ are two Hermitian operators on $\mathcal{H}_{\rm S}$ corresponding to 
the Lagrange multipliers 
associated with  $M_{\vec{p}}\geqslant0$ and $\mathbbm{1}-M_{\vec{p}}\geqslant0$.
As in the primal problem, a pair $(E,K)$ is said to be feasible
if they satisfy the dual constraints of \eqref{eq:dual}.\\
\indent The first and second constraints of \eqref{eq:dual} 
are the positivity conditions of Lagrange multipliers and
the last condition means the {\em Lagrangian stability}, i.e., $\partial \mathcal{L}/\partial M_{\vec{p}}=0$.
The first constraint of \eqref{eq:dual} does not mean that $E$ is positive semidefinite.
The first and last constraints of \eqref{eq:dual} can be expressed as
\begin{equation}\label{eq:onlyk}
\langle\tilde{\psi}_{i}|K|\tilde{\psi}_{i}\rangle\geqslant\eta_{i}\ \ \forall i,
\end{equation}
therefore $E$ can be omitted.
When $E$ is omitted, $K$ is called feasbile
if it is positive-semidefinite and satisfies Inequality \eqref{eq:onlyk} for all $i$.
We use the superscript $^{\star}$ to express the optimality of primal and dual variables.\\
\indent The primal optimal value $P_{\max}$ is 
an lower bound of the dual optimal value ${\rm Tr}K^{\star}$ because
\begin{eqnarray}\label{eq:laginq}
{\rm Tr}(\rho M_{\vec{p}})\leqslant 
\mathcal{L}(M_{\vec{p}},E,K)={\rm Tr}K
\end{eqnarray}
for any feasible $M_{\vec{p}}$ and $(E,K)$.
From Inequality \eqref{eq:laginq}, it can be easily seen that 
feasible $M_{\vec{p}}$ and $(E,K)$ with ${\rm Tr}(\rho M_{\vec{p}})={\rm Tr}K$ satisfy
\begin{equation}\label{eq:comslk}
{\rm Tr}(M_{\vec{p}}E)=
{\rm Tr}[(\mathbbm{1}-M_{\vec{p}})K]=0,
\end{equation}
which is called the {\em complementary slackness condition}.
Applying $K=\rho+E$ to ${\rm Tr}(M_{\vec{p}}E)=0$, we can have
\begin{equation}\label{eq:comcon}
{\rm Tr}[M_{\vec{p}}(K-\rho)]=0
\end{equation}
or equivalently,
\begin{equation}
p_{i}(\langle\tilde{\psi}_{i}|K|\tilde{\psi}_{i}\rangle-\eta_{i})=0 \ \ \forall i
\end{equation}
which are the complementary slackness condition that replaces ${\rm Tr}(M_{\vec{p}}E)=0$ when $E$ is omitted.
Due to the weak duality ${\rm Tr}M_{\vec{p}}^{\star}\leqslant{\rm Tr}K^{\star}$, 
if feasible $M_{\vec{p}}$ and $(E,K)$ satisfying Eq. \eqref{eq:comslk} exist, 
they are optimal and the strong duality ${\rm Tr}M_{\vec{p}}^{\star}={\rm Tr}K^{\star}$ holds.
In addition, when strong duality holds, 
the optimal primal and dual variables 
satisfy Eq. \eqref{eq:comslk}.\\
\indent The convex optimization problem given in \eqref{eq:primal} satisfies 
the well-known {\em Slater's condition}\cite{boyd2004}\,(i.e., the existence of 
strictly feasible $M_{\vec{p}}$) as a sufficient condition for strong duality,
so primal and dual problems provide the same optimal value, i.e., ${\rm Tr}K^{\star}=P_{\max}$.
This implies that Eq. \eqref{eq:comslk}
is a necessary and sufficient condition for feasible $M_{\vec{p}}$ and $(E,K)$
to be optimal. Therefore, $M_{\vec{p}}$ and $(E,K)$ are optimal if and only if they satisfy 
the so-called {\em Karush-Kuhn-Tucker} (KKT) condition\cite{elda2003,elda2004,boyd2004,naka2015}:
\begin{equation}\label{eq:lcp}
\begin{array}{llll}
 \mbox{(i)}& M_{\vec{p}}\geqslant 0,\ \mathbbm{1}-M_{\vec{p}}\geqslant0;\\[2mm]
\mbox{(ii)}& \langle\tilde{\psi}_{i}|E|\tilde{\psi}_{i}\rangle\geqslant0\ \forall i,\ 
K\geqslant 0,\ K=\rho+E;\\[2mm]
\mbox{(iii)}& {\rm Tr}(M_{\vec{p}}E)=0,\  {\rm Tr}[(\mathbbm{1}-M_{\vec{p}})K]=0.
\end{array}
\end{equation}
The approach to finding an optimal solution using KKT condition 
is called the {\em linear complementarity problem} (LCP) approach\cite{bae2013,bae2015}.
The characteristic of LCP is that it considers both primal and dual variables, 
that it is neither the minimum nor the maximum problem
[i.e., there is no objective function],
and that the constraints are KKT condition \eqref{eq:lcp}.\\
\indent From KKT condition \eqref{eq:lcp}, we can see that
both $\mathbbm{1}-M_{\vec{p}}^{\star}$ and $K^{\star}$ are positive-semidefinite, nonzero, and not full rank. 
$M_{\vec{p}}^{\star}$ is also positive-semidefinite and nonzero, but it may be full rank.
Note that if $(M_{\vec{p}},E,K)$ and $(\bar{M}_{\vec{p}},\bar{E},\bar{K})$ satisfy KKT condition,
so do $(M_{\vec{p}},\bar{E},\bar{K})$ and $(\bar{M}_{\vec{p}},E,K)$ because 
both $M_{\vec{p}}$ and $\bar{M}_{\vec{p}}$ are primal optimal and both $(E,K)$ and $(\bar{E},\bar{K})$ are dual optimal.
Furthermore, the optimality condition for $(M_{\vec{p}},E,K)$ 
can be converted to one for $(M_{\vec{p}},K)$ 
by replacing the constraints for $E$ with \eqref{eq:onlyk} and \eqref{eq:comcon}.
Thus, $(M_{\vec{p}},K)$ is optimal if and only if 
\begin{equation}\label{eq:kkt}
\begin{array}{llll}
\mbox{(i)}& p_{i}\geqslant 0\ \forall i,\ \mathbbm{1}-M_{\vec{p}}\geqslant0;\\[2mm]
\mbox{(ii)}& K\geqslant 0,\ {\rm Tr}[(\mathbbm{1}-M_{\vec{p}})K]=0;\\[2mm]
\mbox{(iii)}& p_{i}(\langle\tilde{\psi}_{i}|K|\tilde{\psi}_{i}\rangle-\eta_{i})=0\ \forall i; \\[2mm]
\mbox{(iv)}& \langle\tilde{\psi}_{i}|K|\tilde{\psi}_{i}\rangle\geqslant\eta_{i}\ \forall i.
\end{array}
\end{equation}
If Condition \eqref{eq:lcp} holds for $(M_{\vec{p}},K,E)$, then Condition \eqref{eq:kkt} holds for $(M_{\vec{p}},K)$.
Conversely, if $(M_{\vec{p}},K)$ satisfies Condition \eqref{eq:kkt}, 
$(M_{\vec{p}},K,E)$ with $E=K-\rho$ satisfies Condition \eqref{eq:lcp}.\\
\indent We will selectively use Conditions \eqref{eq:lcp} and \eqref{eq:kkt}
because both have their respective advantages.
Eliminating the dual variable $E$ simplifies  
the form of the dual problem and KKT condition,
so Condition \eqref{eq:kkt} is useful 
for demonstrating the optimality of given variables or 
for deriving the condition that fixed variables must be optimal.
On the other hand, leaving $E$ as in Condition \eqref{eq:lcp} allows a useful condition, Lagrangian stability, to be used.
The Lagrangian stability condition is useful for demonstrating the uniqueness of optimal variables.\\
\indent {\bf Lemma 1.} $\vec{p}^{\star}$ is unique.\\
\indent {\bf Proof.} Assume that both $M_{\vec{p}}$ and $\bar{M}_{\vec{p}}$ are primal optimal and $(K,E)$ is dual optimal.
Then, both $(M_{\vec{p}},E,K)$ and $(\bar{M}_{\vec{p}},E,K)$ satisfy Condition \eqref{eq:lcp}. 
Since ${\rm Tr}[(\mathbbm{1}-M_{\vec{p}})K]=0$, it follows that 
\[M_{\vec{p}}K=KM_{\vec{p}}=K.\]
In a similar way, 
\[\bar{M}_{\vec{p}}K=K\bar{M}_{\vec{p}}=K.\]
\indent From the above equation and $K=\rho+E$, we obtain
\[\begin{array}{ll}
\rho M_{\vec{p}}+E M_{\vec{p}}=
K=\rho \bar{M}_{\vec{p}}+E \bar{M}_{\vec{p}}.
\end{array}\]
From Eq. \eqref{eq:rhomp}, we have
\[
\rho M_{\vec{p}}=\sum_{i=1}^{n}|\psi_{i}\rangle\langle\tilde{\psi}_{i}|(\rho M_{\vec{p}}+E M_{\vec{p}})|\psi_{i}\rangle\langle\tilde{\psi}_{i}|
=\sum_{i=1}^{n}|\psi_{i}\rangle\langle\tilde{\psi}_{i}|(\rho\bar{M}_{\vec{p}}+E\bar{M}_{\vec{p}})|\psi_{i}\rangle\langle\tilde{\psi}_{i}|=\rho\bar{M}_{\vec{p}}
\]
in which the first and last equalities hold due to
\[{\rm Tr}(M_{\vec{p}}E)={\rm Tr}(\bar{M}_{\vec{p}}E)=0.\]
Since $\rho$ is invertible, $M_{\vec{p}}=\bar{M}_{\vec{p}}$.
This means that $M_{\vec{p}}^{\star}$ is unique. Therefore, $p^{\star}$ is unique.\qed\\
\indent The uniqueness of optimal primal variables tells us that 
the optimal UD of $n$ linearly independent pure states is a problem of 
finding only one point $\vec{p}^{\star}$ in $\mathbb{R}^{n}$.\\
\indent The optimal dual variable $K^{\star}$ is generally not unique, but it is uniquely determined
when the zero eigenvalue of $\mathbbm{1}-M_{\vec{p}}^{\star}$ is not degenerate\cite{pang2009}.
In this case, it is easy to find the detail relation between 
$\{\eta_{i},|\psi_{i}\rangle\}_{i=1}^{n}$ and the optimal UD measurement(or $\vec{p}^{\star}$).
The simplest example is the case where only $p_{1}^{\star}$ is nonzero.
In this case, the optimal UD measurement detects only $|\psi_{1}\rangle$ unambiguously.
%%%%%%%%%%%%%%%%(SUBSECTION)
\subsection*{Proof of Theorem 1}
\indent In this section, we prove Theorem 1 using the optimality condition in \eqref{eq:kkt}. For the necessity, suppose that 
\[p_{2}^{\star}=\cdots=p_{n}^{\star}=0.\]
Because of primal constraints (i),
\[0\leqslant p_{1}^{\star}\leqslant |\langle\tilde{\psi}_{1}|\tilde{\psi}_{1}\rangle|^{-1}.\] 
Since the kernel of $\mathbbm{1}-M_{\vec{p}}^{\star}$ must have a nonzero vector, 
\[p_{1}^{\star}=|\langle\tilde{\psi}_{1}|\tilde{\psi}_{1}\rangle|^{-1}.\]
That is, 
\[M_{\vec{p}}^{\star}=|\langle\tilde{\psi}_{1}|\tilde{\psi}_{1}\rangle|^{-1}|\tilde{\psi}_{1}\rangle\langle\tilde{\psi}_{1}|.\]
\indent Because $K^{\star}$ is a positive-semidefinite operator on the kernel of $\mathbbm{1}-M_{\vec{p}}^{\star}$,
\[K^{\star}=a|\tilde{\psi}_{1}\rangle\langle\tilde{\psi}_{1}|\] 
for some nonnegative number $a$. 
By (iii) of Condition \eqref{eq:kkt},
$a$ is determined as $\eta_{1}|\langle\tilde{\psi}_{1}|\tilde{\psi}_{1}\rangle|^{-2}$,
that is,
\[K^{\star}=\eta_{1}|\langle\tilde{\psi}_{1}|\tilde{\psi}_{1}\rangle|^{-2}|\tilde{\psi}_{1}\rangle\langle\tilde{\psi}_{1}|.\]
Due to (iv) of Condition \eqref{eq:kkt}, the following inequality holds,
\begin{equation*}
\frac{\eta_{i}}{\eta_{1}}\leqslant\frac{\langle\tilde{\psi}_{i}|K^{\star}|\tilde{\psi}_{i}\rangle}{\eta_{1}}
=\frac{|\langle\tilde{\psi}_{i}|\tilde{\psi}_{1}\rangle|^{2}}{|\langle\tilde{\psi}_{1}|\tilde{\psi}_{1}\rangle|^{2}}\ \ \forall i,
\end{equation*}
which proves the necessity of our theorem.\\
\indent To prove sufficiency, we assume that 
\[\frac{\eta_{i}}{\eta_{1}}\leqslant\frac{|\langle\tilde{\psi}_{i}|\tilde{\psi}_{1}\rangle|^{2}}{|\langle\tilde{\psi}_{1}|\tilde{\psi}_{1}\rangle|^{2}}\ \ \forall i.\]
It is straightforward to verify that the following $(M_{\vec{p}},K)$ satisfies Condition \eqref{eq:kkt};
\[M_{\vec{p}}=|\langle\tilde{\psi}_{1}|\tilde{\psi}_{1}\rangle|^{-1}|\tilde{\psi}_{1}\rangle\langle\tilde{\psi}_{1}|\]
and
\[K=\eta_{1}|\langle\tilde{\psi}_{1}|\tilde{\psi}_{1}\rangle|^{-2}|\tilde{\psi}_{1}\rangle\langle\tilde{\psi}_{1}|.\]
Thus, $(M_{\vec{p}},K)$ is optimal, so 
\[p_{i}^{\star}=\delta_{i1}|\langle\tilde{\psi}_{1}|\tilde{\psi}_{1}\rangle|^{-1},\]
and $p_{2}^{\star}=\cdots=p_{n}^{\star}=0$. 
This completes our proof.
%%%%%%%%%%%%%%%%(SUBSECTION)
\subsection*{Proof of Theorem 2}
In this section, we suppose that only $|\tilde{\Psi}_{1}\rangle$ is a product vector,
which implies only $|\Psi_{1}\rangle$ can be unambiguously discriminated using LOCC.\\
\indent To prove the necessity, we assume that the globally optimal UD of $\{\xi_{i},|\Psi_{i}\rangle\}_{i=1}^{n}$ is possible using LOCC. Thus, the globally optimal UD of $\{\xi_{i},|\Psi_{i}\rangle\}_{i=1}^{n}$ is 
the optimally unambiguous detection of $|\Psi_{1}\rangle$.
From Theorem 1, Inequality \eqref{eq:onlyonecon} holds for each $i$,
which proves the necessity of our theorem.\\
\indent To prove sufficiency, suppose Inequality \eqref{eq:onlyonecon} holds for each $i$.
According to Theorem 1, the globally optimal UD measurement of $\{\xi_{i},|\Psi_{i}\rangle\}_{i=1}^{n}$ 
detects only one state $|\Psi_{1}\rangle$ unambiguously.
Thus, the optimally unambiguous detection of $|\Psi_{1}\rangle$ means 
the projective measurement 
\[\{|\langle\tilde{\Psi}_{1}|\tilde{\Psi}_{1}\rangle|^{-1}|\tilde{\Psi}_{1}\rangle\langle\tilde{\Psi}_{1}|,\mathbbm{1}-|\langle\tilde{\Psi}_{1}|\tilde{\Psi}_{1}\rangle|^{-1}|\tilde{\Psi}_{1}\rangle\langle\tilde{\Psi}_{1}|\}.\]
Moreover, this projective measurement can be done using finite-round LOCC.
To show this, let $\mathcal{H}_{k}$ be each party's Hilbert space, i.e., $\mathcal{H}=\bigotimes_{k=1}^{m}\mathcal{H}_{k}$.
Since $|\tilde{\Psi}_{1}\rangle$ is a product vector, its normalized form can be written as 
\[\begin{array}{c}
|\langle\tilde{\Psi}_{1}|\tilde{\Psi}_{1}\rangle|^{-1/2}
|\tilde{\Psi}_{1}\rangle=\bigotimes_{k=1}^{m}|\phi_{k}\rangle,
\end{array}\]
in which $|\phi_{k}\rangle$ is a pure state in $\mathcal{H}_{k}$.
Each party performs a projective measurement $\{|\phi_{k}\rangle\langle\phi_{k}|,\mathbbm{1}_{k}-|\phi_{k}\rangle\langle\phi_{k}|\}$, and all parties share their local measurement results through classical communication.
Here $\mathbbm{1}_{k}$ is the identity operator on $\mathcal{H}_{k}$.
If the measurement result is $|\phi_{k}\rangle\langle\phi_{k}|$ for all party $k$, 
they unambiguously discriminate $|\Psi_{1}\rangle$, otherwise
it is considered as inconclusive results.
Thus, we can detect only $|\Psi_{1}\rangle$ unambiguously and optimally using only finite-round LOCC.
This implies that the globally optimal UD of $\{\xi_{i},|\Psi_{i}\rangle\}_{i=1}^{n}$ can be achieved using LOCC,
which proves the sufficiency of our theorem.
%%%%%%%%%%%%%%%%(SUBSECTION)
\subsection*{Nonexistence of complete $2\otimes2$ product basis where only one state can be unambiguously discriminated}
\indent Suppose that $\{|\Psi_{i}\rangle\}_{i=1}^{4}$ is a normalized basis of $\mathbb{C}^{2}\otimes\mathbb{C}^{2}$ and consists of four product states $|\Psi_{i}\rangle=|\alpha_{i}\rangle\otimes|\beta_{i}\rangle$.
Also, assume that the reciprocal vector $|\tilde{\Psi}_{1}\rangle$ is a product vector, i.e., $|\tilde{\Psi}_{1}\rangle=|\tilde{\alpha}\rangle\otimes|\tilde{\beta}\rangle$. This assumption implies that $|\tilde{\alpha}\rangle$ is orthogonal to two of $\{|\alpha_{i}\rangle\}_{i=2}^{4}$ or $|\tilde{\beta}\rangle$ is orthogonal to two of $\{|\beta_{i}\rangle\}_{i=2}^{4}$.\\
\indent When $|\alpha_{2}\rangle$ and $|\alpha_{3}\rangle$ are orthogonal to $|\tilde{\alpha}\rangle$,
the two vectors are linearly dependent because they are in $\mathbb{C}^{2}$.
Since $|\Psi_{2}\rangle$ and $|\Psi_{3}\rangle$ are linearly independent,
$|\beta_{2}\rangle$ and $|\beta_{3}\rangle$ span $\mathbb{C}^{2}$,
and thus all vectors orthogonal to both $|\Psi_{2}\rangle$ and $|\Psi_{3}\rangle$ are in
$\{|\tilde{\alpha}\rangle\otimes|\varphi\rangle:|\varphi\rangle\in\mathbb{C}^{2}\}$.
This means that $|\tilde{\Psi}_{4}\rangle$ is a product vector because $\langle\tilde{\Psi}_{4}|\Psi_{i}\rangle=0$ for $i\neq 4$.
In any case, we can similarly show that
at least one of $\{|\tilde{\Psi}_{i}\rangle\}_{i=2}^{4}$ is a product vector.
Therefore, there is no set of four linearly independent $2\otimes2$ product states where
only one state can be unambiguously discriminated.
%%%%%%%%%%%%%%%% Acknowledgements %%%%%%%%%%%%%%%
\section*{Acknowledgements}
This work was supported by Quantum Computing Technology Development Program(NRF-2020M3E4A1080088) through the National Research Foundation of Korea(NRF) grant funded by the Korea government(Ministry of Science and ICT).
%%%%%%%%%%%%%%%% References %%%%%%%%%%%%%%%%


\begin{thebibliography}{61}
\bibitem{alocc} LOCC consists of finite- and infinite-round LOCC. Asymptotic LOCC is the closure of finite-round LOCC. It contains LOCC. We also note that SEP contains asymptotic LOCC, but not vice versa.
\end{thebibliography}
\end{document}